\documentclass[aps,prl,showpacs,preprintnumbers,amsmath,amssymb]{revtex4}
\usepackage{setspace}
\usepackage{graphicx}
\begin{document}

\title {A General Method for Deriving Vector Potentials Produced by 
Knotted Solenoids} 

\author{V. V. Sreedhar\footnote{\sl sreedhar@cmi.ac.in}}  

\affiliation{Chennai Mathematical Institute, Plot H1, SIPCOT IT 
Park, Siruseri, Kelambakkam Post, Chennai 603103, India}

\begin{abstract}
A general method for deriving exact expressions for vector potentials
produced by arbitrarily knotted solenoids is presented. It consists 
of using simple physics ideas from magnetostatics to evaluate the 
magnetic field in a surrogate problem. The latter is obtained by  
modelling the knot with wire segments carrying steady currents on a 
cubical lattice. The expressions for a $3_1$ (trefoil) and a $4_1$ 
(figure-eight) knot are explicitly worked out. The results are of some 
importance in the study of the Aharonov-Bohm effect generalised to a 
situation in which charged particles moving through force-free regions 
are scattered by fluxes confined to the interior of knotted impenetrable 
tubes.  

\end{abstract}


\maketitle

The physical importance of the vector potential experienced by a
charged particle travelling through force-free regions was first
stressed in a celebrated paper by Aharonov and Bohm \cite{ab}; although 
it was anticipated much earlier by Ehrenberg and Siday \cite{es}. Landmark
experiments by Chambers \cite{chambers} and later by Tonomura {\it et al}
 \cite{tono} established the results by studying the shifts in the electron 
interference patterns produced by solenoidal and toroidal windings 
respectively. It is expected that in the foreseeable future, these 
experiments may be repeated with more complicated current distributions, 
namely knotted solenoids \cite{abknot}\cite{buniy}.    

In a recent paper the author has presented an exact expression for 
a flat connection on the complement of a torus knot \cite{vvs}. This 
was accomplished by computing the vector potential produced by an 
infinitesimally thin knotted solenoid which effectively confines the 
magnetic flux to the interior of an $\epsilon$-neighbourhood of the knot. 
The choice of toroidal coordinates greatly facilitated the calculation
since arbitrary torus knots can be regarded as closed loops winding, 
a relatively prime number of times, around the two inequivalent cycles 
of a putative torus. 
 
Although the ideas of [7] were born out of a different motivation, they 
are fairly general. Besides, the choice of toroidal coordinates limits 
their applicability to torus knots. To apply the ideas fruitfully to more 
general knots we need to develop a different technique. Towards this end, 
we replace the putative torus mentioned above by a putative cubical lattice. 
The knot is then defined as a self-avoiding polygon with vertices at 
prescribed lattice sites. This way of defining a knot has several advantages. 
First, it is tailor-made for testing important results using computers. 
Second, the lattice spacing introduces a natural cut-off which allows us to 
distinguish between physical knots (which have a finite thickness) and 
mathematical knots (which have vanishing thickness) \cite {hayes}. Third, 
it can be used to discuss any knot -- not just a torus knot. Finally, it 
allows us to use Cartesian coordinates. 

In what follows we use the aforementioned motivations to derive a general 
expression for the vector potential produced in the force-free region
by a solenoidal winding around an arbitrary knot on a cubical lattice. 
As special cases, we then explicitly work out the exact expressions for 
the $3_1$ (trefoil) and the $4_1$ (figure-eight) knots. In principle this 
is a straightforward exercise and consists in putting together the 
contributions to the vector potential coming from the various finite-length
segments of the polygonal knot on the cubical lattice. As is well-known,
the vector potential produced by a finite-length solenoid is given by 
a complicated expression involving elliptic integrals \cite{callaghan}.
So putting together such expressions coming from all the line segments
is an arduous task.

The paper relies on the following thesis: Given a knot, consider a small 
$\epsilon$-neighbourhood around it. Imagine a closely-spaced winding of a 
wire carrying uniform current around this tube. In the limit of small radius 
of cross-section of the tube, the current distribution reduces to a collection 
of magnetic dipoles (tiny magnets) which are lined up along the knot. Since 
the vector potential produced by a magnetic dipole at a given point is given 
by a well-known formula, the coveted answer for the vector potential produced 
by the knot is obtained by integrating the dipole result over the length of 
the knot. This expression has the Biot-Savart form. As already mentioned in 
[7], this is a consequence of the fact that, in the magnetostatic limit, 
the magnetic field due to a current wire and the vector potential due to 
an infinitesimally thin solenoid of the same shape and size satisfy the 
same equations and hence have the same solutions in the region of interest
{\it viz.} the source-free knot complement. In view of this, it is sufficient 
to solve the surrogate problem of finding the magnetic field produced by a 
steady filamentary current, in the shape of the knot under consideration, and 
to formally identify the answer with the vector potential produced by the 
corresponding knot tube. The holonomy viz. the line integral of the vector 
potential around a closed loop in the latter problem is then formally equal to 
the line integral of the magnetic field along the same closed loop in 
the surrogate problem which, as a consequence of Stokes' law, equals the 
current passing through the knot and measures the linking number of the 
loop with the knot under consideration. In other words, the role of the flux 
in the original problem is played by the current in the surrogate problem.

We note that, on the cubical lattice, a polygonal knot (tube) is modelled by 
a collection of finite-length solenoids. Hence, solving the surrogate problem 
amounts to putting together contributions coming from finite-length wire 
segments which make up the knot, each of which has its own length and is 
parallel to one of the coordinate axes.  

A lattice realisation for an arbitrary knot can be obtained by starting at a 
given site on a cubical lattice and setting up a self-avoiding random walk. 
This allows a hypothetical particle at a given site to move to a neighbouring 
site at random, so long as such a site was hitherto not visited; till it 
finally returns to the starting point, which is the only site that can 
be visited twice \cite{hayes}. The (closed) trajectory of the particle 
then describes the knot. 

A more systematic and instructive way of 
constructing lattice knots comes from the connection between knots and 
braids \cite{alexander}. Consider, for example, the trefoil knot. As is 
well-known, it can be obtained by end-on-end closure of the element 
$(\sigma_1)^3$ in ${\hbox{\bf B}}_2$ -- the braid group on two strings. 
Now, instead of looking at the two-dimensional projection of the braid, 
suppose we look at it from the side so that we can perceive the depth. 
From this perspective, over-passes and under-passes are not represented 
by continuous and broken arcs, but manifest themselves as windings between 
the strings which are separated by a distance. Flatten out the arcs into 
straight line segments and imagine that one string lies on the front face 
and the other on the back face of a putative cubical lattice. A crossing 
corresponds to the strings migrating from one face to the other. Now 
choose a grid such that the ends of the line segments lie on lattice sites.
Finally, join the loose ends of the strings along trivial paths (those which
don't introduce new crossings) on the lattice and put a bounding box to 
get a lattice trefoil knot. The lattice figure-eight knot can similarly be 
obtained by the closure of the element $\sigma_1 \sigma_2^{-1}\sigma_1 
\sigma_2^{-1}$ in ${\hbox{\bf B}}_3$ -- the braid group on three strings. 

Clearly there are many ways to produce a lattice version of a knot by the 
above presciption. The most economical way is to minimise the stick number 
(number of line segments) and the step number (related to the size of the 
lattice) of the knot. Important theorems regarding the minimum stick number and
minimum step number for popular knots have been proved recently \cite{huhoh}. 
The figures for $3_1$ and $4_1$ knots with minimum stick numbers, adapted 
from \cite{huhoh} are shown below.    

\begin{widetext}
\begin{figure} [<h>]
\includegraphics[scale=0.7]{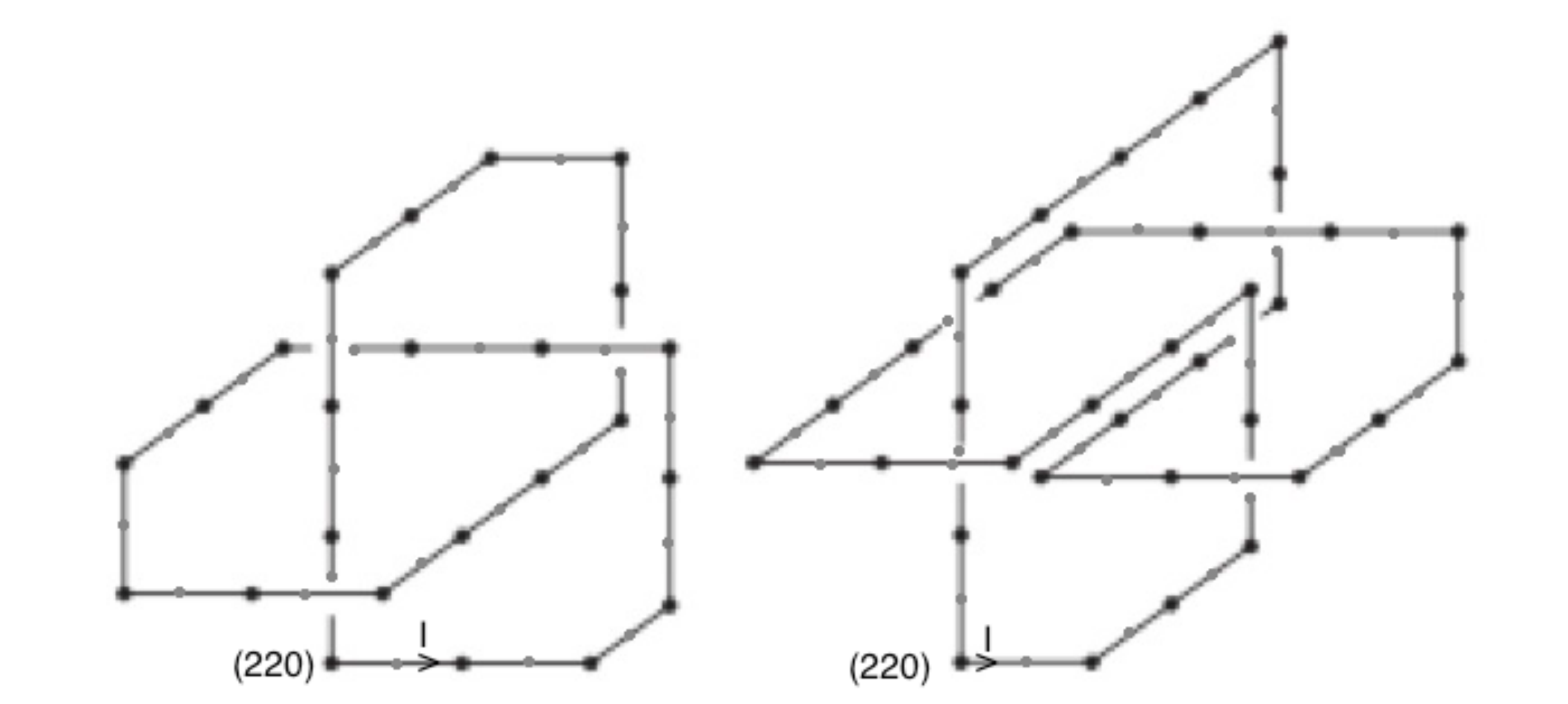}

\caption{$3_1$ (Trefoil) and $4_1$ (Figure-Eight) Lattice Knots with Minimum 
Stick Numbers}
\label{latknot}
\end{figure}
\end{widetext}

Taking (220) as the fiducial point, we label the successive line segments in 
the direction of the current flow, by the letter $\alpha$, starting with 
$\alpha =1$. Clearly, for the trefoil, $\alpha = 1,2,\cdots 12$ and for the 
figure-eight knot, $\alpha = 1,2,\cdots 14$. We double the number of sites 
from the minimum suggested by the diagrams. This is indicated by the lighter
dots in the pictures. While this will not increase the 
complexity of the computation in any way, it will be useful subsequently in
checking the non-triviality of the flat connection. We denote by $I$, the 
uniform current passing through the circuits. 

Before we compute the magnetic field $\vec B$, due to a polygonal knot 
carrying uniform current $I$, at any point $P$ with coordinates $\vec r$, 
it is useful to recall the corresponding result for a finite straight wire
carrying a steady current. As is well-known \cite {tipler}, this is 
obtained by a straightforward application of the Biot-Savart law, and is 
given by following expression for a thin straight wire placed along
the $x$-axis:   
\begin{equation}
\vec B (\vec r) =  (\frac {I}{c})(\frac{\hat\phi}{r})
({\hbox {cos}} \theta_1 + {\hbox{cos}}\theta_2)  
\end{equation}
where $r$ is the perpendicular distance of the point $P$ from the wire, 
$\theta_1$ and $\theta_2$ are the interior angles subtended by the 
line joining the point $P$ to the extremities of the wire and $\hat\phi$
is the unit vector in the azimuthal direction defined by the right-hand 
rule in the plane perpendicular to the wire. The field 
produced by the knot can now be written down easily by adding the contributions
from the finite-length straight wire segments that make up the knot and 
is given by   
\begin{equation}
\vec B (\vec r) = \sum_\alpha\vec B_\alpha~~~~{\hbox{where}}~~~~
\vec B_\alpha = (\frac {I}{c})(\frac{\hat\phi_{\alpha}}{r_{\alpha}})
({\hbox {cos}} \theta^\alpha_1 + {\hbox{cos}}\theta^\alpha_2)  
\end{equation}
In the above expression, we use the obvious notations: $\vec B_\alpha$ is 
the field produced by the $\alpha$-th segment,  $r_\alpha$ is the 
perpendicular distance from the point $P$ to the segment $\alpha$,  
$\theta^\alpha_1$ and $\theta^\alpha_2$ are the interior angles subtended 
by the lines joining the point $P$ to the extremities of the segment 
$\alpha$, and $\hat\phi_\alpha$ is the unit vector in the azimuthal 
direction in the plane perpendicular to the segment $\alpha$ and is 
given by the right-hand rule. For example, if the segment is parallel 
to the $z$-axis, $\hat \phi = -y\hat i + x\hat j$, {\it etc.} Since a 
general knot will have segments parallel to all the three axes, the total 
magnetic field will have non-vanishing contributions along all the three 
Cartesian basis vectors. In view of this, we can re-express 
the magnetic field in (2) as follows
\begin{equation}
\vec B (x,y,z) = \sum_{i=1}^3 B_i\hat e_i
\end{equation}
Comparing the above two equations, we get the following expression for
the Cartesian components of the magnetic field 
\begin{equation}
B_i = \frac {I}{c} \sum_\alpha \gamma_{\alpha i} 
\frac {({\hbox {cos}} \theta^\alpha_1 + {\hbox{cos}}\theta^\alpha_2)}  
{r_\alpha}
\end{equation}
where the weights $\gamma_{\alpha i}$ for a given $\alpha$ are given by 
$\gamma_{\alpha i} = \hat\phi_\alpha\cdot \hat e_i$.

Given a knot it is straightforward to evaluate the $\gamma_{\alpha i}$.
As already mentioned, one may formally identify these expressions for 
the magnetic field produced by a knotted current wire with the expressions
for the vector potential produced by solenoidal windings around the 
corresponding knots. 

Substituting the $\gamma_{\alpha i}$ and summing over $\alpha$ we get, 
after suppressing a factor of $I/c$ in all the equations, the following 
expressions for the magnetic field of the $3_1$ knot:
\begin{widetext}
\begin{equation}
\begin{split}
B_x =& 
-\frac{z}{(6-x)^2 + z^2}[\frac{4-y}{\sqrt {(6-x)^2+ (4-y)^2 + z^2}}
-{2-y\over\sqrt {(6-x)^2 + (2-y)^2 + z^2}}]\cr
&-{(4-y)\over (6-x)^2 + (4-y)^2}[{4-z\over \sqrt {(6-x)^2+ (4-y)^2 + (4-z)^2}}
-{-z\over\sqrt {(6-x)^2 + (4-y)^2 + z^2}}]\cr
&+{(4-z)\over x^2 + (4-z)^2} [{-y\over \sqrt {x^2+ y^2 + (4-z)^2}}
- {(4-y)\over\sqrt {x^2 + (4-y)^2 + (4-z)^2}}]\cr
& +{(2-z)\over (4-x)^2 + (2-z)^2}[{(6-y)\over\sqrt {(4-x)^2+ (6-y)^2 + (2-z)^2}}
-{(-y)\over\sqrt {(4-x)^2 + y^2 + (2-z)^2}}]\cr
& -{(6-y)\over (4-x)^2 + (6-y)^2}[{(6-z)\over\sqrt {(4-x)^2+ (6-y)^2 + (6-z)^2}}
-{(2-z)\over\sqrt {(4-x)^2 + (6-y)^2 + (2-z)^2}}]\cr
& +{(6-z)\over (2-x)^2 + (6-z)^2}[{(2-y)\over\sqrt {(2-x)^2+ (2-y)^2 + (6-z)^2}}
-{(6-y)\over\sqrt {(2-x)^2 + (6-y)^2 + (6-z)^2}}]\cr
& -{(2-y)\over (2-x)^2 + (2-y)^2}[{(-z)\over\sqrt {(2-x)^2+ (2-y)^2 + z^2}}
-{(6-z)\over\sqrt {(2-x)^2 + (2-y)^2 + (6-z)^2}}]\cr
& +{y\over x^2 + y^2}[{(2-z)\over\sqrt {x^2+ y^2 + (2-z)^2}}
-{(4-z)\over\sqrt {x^2 + y^2 + (4-z)^2}}]\cr
\end{split}
\end{equation}
\begin{equation}
\begin{split}
B_y =& 
\frac{z}{(2-y)^2 + z^2}[\frac{6-x}{\sqrt {(6-x)^2+ (2-y)^2 + z^2}}
-{2-x\over\sqrt {(2-x)^2 + (2-y)^2 + z^2}}]\cr
&+ {(6-x)\over (6-x)^2 + (4-y)^2}[{4-z\over \sqrt {(6-x)^2+ (4-y)^2 + (4-z)^2}}
-{-z\over\sqrt {(6-x)^2 + (4-y)^2 + z^2}}]\cr
&-{(4-z)\over (4-y)^2 + (4-z)^2} [{6-x\over \sqrt {(6-x)^2+ (4-y)^2 + (4-z)^2}}
- {-x)\over\sqrt {x^2 + (4-y)^2 + (4-z)^2}}]\cr
& {-x\over x^2 + y^2}[{(2-z)\over\sqrt {x^2 + y^2 + (2-z)^2}}
-{(4-z)\over\sqrt {x^2 + y^2 + (4-z)^2}}]\cr
& -{(2-z)\over y^2 + (2-z)^2}[{(4-x)\over\sqrt {(4-x)^2+ y^2 + (2-z)^2}}
-{-x\over\sqrt {x^2 + y^2 + (2-z)^2}}]\cr
& +{(4-x)\over (4-x)^2 + (6-y)^2}[{(6-z)\over\sqrt {(4-x)^2+ (6-y)^2 + (6-z)^2}}
-{(2-z)\over\sqrt {(4-x)^2 + (6-y)^2 + (2-z)^2}}]\cr
& -{(6-z)\over (6-y)^2 + (6-z)^2}[{(2-x)\over\sqrt {(2-x)^2+ (6-y)^2 + (6-z)^2}}
-{(4-x)\over\sqrt {(4-x)^2 + (6-y)^2 + (6-z)^2}}]\cr
& +{(2-x)\over (2-x)^2 + (2-y)^2}[{-z\over\sqrt {(2-x)^2+ (2-y)^2 + z^2}}
-{(6-z)\over\sqrt {(2-x)^2 + (2-y)^2 + (6-z)^2}}]\cr
\end{split}
\end{equation}
\begin{equation}
\begin{split}
B_z =& 
\frac{(2-y)}{(2-y)^2 + z^2}[\frac{6-x}{\sqrt {(6-x)^2+ (2-y)^2 + z^2}}
-{2-x\over\sqrt {(2-x)^2 + (2-y)^2 + z^2}}]\cr
&- {(6-x)\over (6-x)^2 + z^2}[{4-y\over \sqrt {(6-x)^2+ (4-y)^2 + z^2}}
-{-(2-y)\over\sqrt {(6-x)^2 + (2-y)^2 + z^2}}]\cr
&{(4-y)\over (4-y)^2 + (4-z)^2} [{6-x\over \sqrt {(6-x)^2+ (4-y)^2 + (4-z)^2}}
- {-x)\over\sqrt {x^2 + (4-y)^2 + (4-z)^2}}]\cr
& {x\over x^2 + (4-z)^2}[{-y\over\sqrt {x^2 + y^2 + (4-z)^2}}
-{(4-y)\over\sqrt {x^2 + (4-y)^2 + (4-z)^2}}]\cr
& -{y\over y^2 + (2-z)^2}[{(4-x)\over\sqrt {(4-x)^2+ y^2 + (2-z)^2}}
-{-x\over\sqrt {x^2 + y^2 + (2-z)^2}}]\cr
& -{(4-x)\over (4-x)^2 + (2-z)^2}[{(6-y)\over\sqrt {(4-x)^2+ (6-y)^2 + (2-z)^2}}
-{-y\over\sqrt {(4-x)^2 + y^2 + (2-z)^2}}]\cr
& +{(6-y)\over (6-y)^2 + (6-z)^2}[{(2-x)\over\sqrt {(2-x)^2+ (6-y)^2 + (6-z)^2}}
-{(4-x)\over\sqrt {(4-x)^2 + (6-y)^2 + (6-z)^2}}]\cr
& -{(2-x)\over (2-x)^2 + (6-z)^2}[{(2-y)\over\sqrt {(2-x)^2+ (2-y)^2 + (6-z)^2}}
-{(6-y)\over\sqrt {(2-x)^2 + (6-y)^2 + (6-z)^2}}]\cr
\end{split}
\end{equation}
\end{widetext}
The corresponding expressions for the $4_1$ (figure-eight) knot can be
obtained exactly in the same manner and are given below. 
\begin{widetext}
\begin{equation}
\begin{split}
B_x =& 
-\frac{z}{(4-x)^2 + z^2}[\frac{6-y}{\sqrt {(4-x)^2+ (6-y)^2 + z^2}}
-{2-y\over\sqrt {(4-x)^2 + (2-y)^2 + z^2}}]\cr
&-{(6-y)\over (4-x)^2 + (6-y)^2}[{4-z\over \sqrt {(4-x)^2+ (6-y)^2 + (4-z)^2}}
-{-z\over\sqrt {(4-x)^2 + (6-y)^2 + z^2}}]\cr
&+{(4-z)\over (4-x)^2 + (4-z)^2} [{-y\over \sqrt {(4-x)^2+ y^2 + (4-z)^2}}
- {(6-y)\over\sqrt {(4-x)^2 + (6-y)^2 + (4-z)^2}}]\cr
& +{(4-z)\over x^2 + (4-z)^2}[{(8-y)\over\sqrt {x^2+ (8-y)^2 + (4-z)^2}}
-{(-y)\over\sqrt {x^2 + y^2 + (4-z)^2}}]\cr
& -{(8-y)\over (6-x)^2 + (8-y)^2}[{(2-z)\over\sqrt {(6-x)^2+ (8-y)^2 + (2-z)^2}}
-{(4-z)\over\sqrt {(6-x)^2 + (8-y)^2 + (4-z)^2}}]\cr
& +{(2-z)\over (6-x)^2 + (2-z)^2}[{(8-y)\over\sqrt {(6-x)^2+ (8-y)^2 + (2-z)^2}}
-{(4-y)\over\sqrt {(6-x)^2 + (4-y)^2 + (2-z)^2}}]\cr
& +{(2-z)\over (2-x)^2 + (2-z)^2}[{(8-y)\over\sqrt {(2-x)^2+ (8-y)^2 + (2-z)^2}}
-{(4-y)\over\sqrt {(2-x)^2 + (4-y)^2 + (2-z)^2}}]\cr
& -{(10-y)\over (2-x)^2 + (10-y)^2}[{(6-z)\over\sqrt {(2-x)^2+ (10-y)^2 + 
(6-z)^2}}
-{(2-z)\over\sqrt {(2-x)^2 + (10-y)^2 + (2-z)^2}}]\cr
& +{(6-z)\over (2-x)^2 + (6-z)^2}[{(10-y)\over\sqrt {(2-x)^2+ (10-y)^2 
+ (6-z)^2}} -{(2-y)\over\sqrt {(2-x)^2 + (2-y)^2 + (6-z)^2}}]\cr
& -{(2-y)\over (2-x)^2 + (2-y)^2}[{-z\over\sqrt {(2-x)^2+ (2-y)^2 + 
z^2}}
-{(6-z)\over\sqrt {(2-x)^2 + (2-y)^2 + (6-z)^2}}]\cr
\end{split}
\end{equation}
\begin{equation}
\begin{split}
B_y =& 
-\frac{z}{(2-y)^2 + z^2}[\frac{4-x}{\sqrt {(4-x)^2+ (2-y)^2 + z^2}}
-{2-x\over\sqrt {(2-x)^2 + (2-y)^2 + z^2}}]\cr
&+ {(4-x)\over (4-x)^2 + (6-y)^2}[{4-z\over \sqrt {(4-x)^2+ (6-y)^2 + (4-z)^2}}
-{-z\over\sqrt {(4-x)^2 + (6-y)^2 + z^2}}]\cr
&-{(4-z)\over y^2 + (4-z)^2} [{-x\over \sqrt {x^2+ y^2 + (4-z)^2}}
- {(4-x)\over\sqrt {(4-x)^2 + y^2 + (4-z)^2}}]\cr
& {-(4-z)\over (8-y)^2 + (4-z)^2}[{(6-x)\over\sqrt {(6-x)^2 + (8-y)^2 
+ (4-z)^2}}
-{-x\over\sqrt {x^2 + (8-y)^2 + (4-z)^2}}]\cr
& +{(6-x)\over (6-x)^2 + (8-y)^2}[{(2-z)\over\sqrt {(6-x)^2+ (8-y)^2 + (2-z)^2}}
-{(4-z)\over\sqrt {(6-x)^2 + (8-y)^2 + (4-z)^2}}]\cr
& -{(2-z)\over (4-y)^2 + (2-z)^2}[{(2-x)\over\sqrt {(2-x)^2+ (4-y)^2 + (2-z)^2}}
-{(6-x)\over\sqrt {(6-x)^2 + (4-y)^2 + (2-z)^2}}]\cr
& {(2-x)\over (2-x)^2 + (10-y)^2}[{(6-z)\over\sqrt {(2-x)^2+ (10-y)^2 + (6-z)^2}}
-{(2-z)\over\sqrt {(2-x)^2 + (10-y)^2 + (2-z)^2}}]\cr
& +{(2-x)\over (2-x)^2 + (2-y)^2}[{(6-z)\over\sqrt {(2-x)^2+ (2-y)^2 + (6-z)^2}}
-{(2-z)\over\sqrt {(2-x)^2 + (2-y)^2 + (2-z)^2}}]\cr
\end{split}
\end{equation}
\begin{equation}
\begin{split}
B_z =& 
\frac{(2-y)}{(2-y)^2 + z^2}[\frac{4-x}{\sqrt {(4-x)^2+ (2-y)^2 + z^2}}
-{2-x\over\sqrt {(2-x)^2 + (2-y)^2 + z^2}}]\cr
&- {(4-x)\over (4-x)^2 + z^2}[{6-y\over \sqrt {(4-x)^2+ (6-y)^2 + z^2}}
-{(2-y)\over\sqrt {(6-x)^2 + (2-y)^2 + z^2}}]\cr
&-{(4-x)\over (4-x)^2 + (4-z)^2} [{-y\over \sqrt {(4-x)^2+ y^2 + (4-z)^2}}
- {(6-y)\over\sqrt {(4-x)^2 + (6-y)^2 + (4-z)^2}}]\cr
& -{y\over y^2 + (4-z)^2}[{-x\over\sqrt {x^2 + y^2 + (4-z)^2}}
-{(4-x)\over\sqrt {(4-x)^2 + y^2 + (4-z)^2}}]\cr
&+ {x\over x^2 + (4-z)^2}[{(8-y)\over\sqrt {x^2+ (8-y)^2 + (4-z)^2}}
-{-y\over\sqrt {x^2 + y^2 + (4-z)^2}}]\cr
& +{(8-y)\over (8-y)^2 + (4-z)^2}[{(6-x)\over\sqrt {(6-x)^2+ (8-y)^2 + (4-z)^2}}
-{-x\over\sqrt {x^2 + (8-y)^2 + (4-z)^2}}]\cr
& -{(6-x)\over (6-x)^2 + (2-z)^2}[{(4-y)\over\sqrt {(6-x)^2+ (4-y)^2 + (2-z)^2}}
-{(8-y)\over\sqrt {(6-x)^2 + (8-y)^2 + (2-z)^2}}]\cr
& +{(4-y)\over (4-y)^2 + (2-z)^2}[{(2-x)\over\sqrt {(2-x)^2+ (4-y)^2 + (2-z)^2}}
-{(6-x)\over\sqrt {(6-x)^2 + (4-y)^2 + (2-z)^2}}]\cr
& -{(2-x)\over (2-x)^2 + (2-z)^2}[{(8-y)\over\sqrt {(2-x)^2+ (8-y)^2 + (2-z)^2}}
-{(4-y)\over\sqrt {(2-x)^2 + (4-y)^2 + (2-z)^2}}]\cr
& -{(2-x)\over (2-x)^2 + (6-z)^2}[{(2-y)\over\sqrt {(2-x)^2+ (2-y)^2 + (6-z)^2}}
-{(10-y)\over\sqrt {(2-x)^2 + (10-y)^2 + (6-z)^2}}]\cr
\end{split}
\end{equation}
\end{widetext}

It is obvious that the resulting expressions are cumbersome and not very 
illuminating. Hence, we turn our attention towards establishing the 
non-triviality of the connection by examining a holonomy. It is obtained 
by evaluating the line integral of the connection for the $3_1$ knot above 
along the closed path $(532) - (732) - (752) - (552) - (532)$. This path lies 
in the $xy$-plane. We expect the answer to furnish the value of the flux 
through the surface bounded by the loop, which in turn is the flux contained 
in the segment $\alpha = 3$, which is parallel to the $z$-axis, and passes 
through the centre $(321)$ of the surface. In the surrogate problem this is 
just the current passing through the segment $\alpha =3$. A painstaking 
sanity-check, using Mathematica, reveals that the answer for the holonomy is 
indeed what is expected. Similar tests can be carried out with various 
choices of the holonomy both for the $3_1$ and $4_1$ knots. 

To summarise, we have worked out explicit expressions for the vector 
potentials produced by knotted solenoids. This was accomplished by 
working out the expressions for the magnetic fields produced by
uniform currents running through corresponding knotted wire segments. 
The answers we obtain are non-trivial by construction, because they follow 
from a nontrivial solution of Maxwell's equations, namely, Biot-Savart law, 
for a given current distribution. The non-triviality of the potential is 
further directly verified by explicitly computing certain holonomies. 

It is obvious that the technique presented in this paper is sufficiently
general to be applied for an arbitrary knot. Besides being of interest 
in view of the fact that it yields exact expressions, it is hoped that 
it will be of some use in solving potential problems of the Aharonov-Bohm 
type \cite{ab} and its generalizations \cite{abknot}\cite{buniy}.   

\begin{acknowledgements}

I thank K. G. Arun, G. Krishnaswami, R. Nityananda, T. R. Ramadas, and M. K.
Vemuri for discussions. I also thank an anonymous referee for suggestions 
to improve the presentation. 

\end{acknowledgements}

\end{document}